# Using Binary File Format Description Languages for Documenting, Parsing, and Verifying Raw Data in TAIGA Experiment

1st Bychkov I.[1,2], 2nd Demichev A.[3], 3rd Dubenskaya J.[3], 4th Fedorov O.[4], |
5th Hmelnov A.[1,2], 6th Kazarina Y.[4], 7th Korosteleva E.[3], 8th Kostunin D.[5],
9th Kryukov A.[3], 10th Mikhailov A.[1, a], 11th Nguyen M.D.[3], 12th Polyakov S.[3],
13th Postnikov E.[3], 14th Shigarov A.[1,2,b], 15th Shipilov D.[4], 16th Zhurov D.[4]

[1] *Matrosov Institute for System Dynamics and Control Theory, SB RAS, Irkutsk, Russia*
[2] *Irkutsk State University, Irkutsk, Russia*
[3] *Skobeltsyn Institute of Nuclear Physics, Lomonosov Moscow State University, Moscow, Russia*
[4] *Applied Physics Institute, Irkutsk State University, Irkutsk, Russia*
[5] *Institute for Nuclear Physics, Karlsruhe Institute of Technology, Karlsruhe Germany*

E-mail: [a] mikhailov@icc.ru, [b] shigarov@icc.ru

The paper is devoted to the issues of raw binary data documenting, parsing and verifying in astroparticle data lifecycle. The long-term preservation of raw data of astroparticle experiments as originally generated is essential for re-running analyses and reproducing research results. The selected high-quality raw data should have detailed documentation and accompanied by open software tools for access to them. We consider applicability of binary file format description languages to specify, parse and verify raw data of the Tunka Advanced Instrument for cosmic rays and Gamma Astronomy (TAIGA) experiment. The formal specifications are implemented for five data formats of the experiment and provides automatic generation of source code for data reading libraries in target programming languages (e.g. C++, Java, and Python). These libraries were tested on TAIGA data. They showed a good performance and help us to locate the parts with corrupted data. The format specifications can be used as metadata for exchanging of astroparticle raw data. They can also simplify software development for data aggregation from various sources for the multi-messenger analysis.

# 1. Introduction

The current trend in science is that the researchers from all over the world can immediately get access to research data as soon as they are published. An important topic for modern science in general and astroparticle physics in particular is open science, the model of free access to data (e.g. [1]). It declares that scientific data should be accessible not solely to collaboration members but to all levels of an inquiring society. This approach is especially important in the age of Big Data, when a complete analysis of the experimental data cannot be performed within one collaboration.

Some experiments in astroparticle physics have already adopted this fascinating idea. They have involved their scientific data in electronic publishing, such as KCDC (KASCADE Cosmic ray Data Centre) [2]. KCDC is a web portal where KASCADE-Grande [3] scientific data are made available for the interested public. KCDC is driven within KASCADE-Grande experiment, which is already dismantled. However, TAIGA [4], an operating experiment in Russia, is producing the data for more than ten years. Obviously, various activities should be performed continuously across all stages of the data life cycle in these experiments: collection and storage of data, its processing and analysis, refining the physical model, publication and share, as well as archive and reuse of the data in the future.

One of the important issues is how to efficiently curate raw binary data to support their availability and reuse in future. TAIGA has five unique binary file formats for representing raw data: TAIGA-IACT, TUNKA-HiSCORE [5], TUNKA-133 [5], TUNKA-GRANDE [6], and TUNKA-REX [7]. The long-term preservation of raw binary data as originally generated is essential for re-running analyses and reproducing research results. To be accessible for the scientific community the raw data should be documented in details and accompanied by open and freely available software for accessing to these data. The neglect of this issues may lead to the need for a reverse engineering of their formats.

The state-of-the-art toolsets for formal describing binary data formats provide a satisfactory solution for the issues of raw data documenting, parsing and verifying. This work demonstrates applicability of binary file format description languages to specify, parse and verify raw data of TAIGA experiment. The formal specifications implemented for five formats of the experiment gives possibility for automatic generation of source code of data accessing libraries in target programming languages (e.g. C++, Java, and Python). These libraries were tested on real data. They demonstrated a good performance and helped us to locate files with corrupted data. This result shows ways for describing binary file formats for astroparticle raw binary data share and reuse. It can be interested in other experiments where raw binary data formats remain weakly documented or some parsing libraries for contemporary programming languages are required.

# 2. Binary Data Format Description Languages

There are several alternatives for formal specification of a binary data format. Some of them allow one to generate program libraries for reading binary data in specified formats. Here we consider some of them to choose ones for describing raw data of astroparticle experiments.

Some tools for specifying network protocols (e.g. NETPDL [8], NETPDLFLTR [9] and BINPAC [10]) can serve for describing binary file formats. Since the nature of network protocols requires the sequential reading of data, these languages are appropriate only for formats with a sequential form of information storage (i.e. without pointers). Some of them provide also a program code generation for processing of binary data. NETPDLFLTR generates binary code for network packet filtering. BINPAC allows one generating C++ code for reading packets. The listed tools are very specialized for the network protocols. HUDDL [11], an XML-based language, serves for specifying hydrographic data formats. It is intended to describe streams of binary data. HUDDL specifications can be used for generating a source code in C, C++, and Python language for data reading. Typically, hydrological files contain some sequences of measurements. Therefore, the generated code performs a sequential reading of the data blocks.

The parser generators, namely ANTLR [12] and BISON [13], can be used for automatically building of program code for data reading. However, the use of a binary file format specification as a grammar imposes severe limitations on its capabilities. The parser generators require presenting a binary file format specification as a grammar with fragments in a target language. As a result, the specification is not declarative. In DATASCRIPT [14] data format specifications are used to generate libraries for reading data in Java language. Instead of pointers, DATASCRIPT uses labels that contain expressions with file fragment addresses. A specification is considered as a set of data type definitions. A separate set of simple bit data types (bit fields) allows one describing bit-oriented data. However, the project has not been updated since 2003 [15].

FLEXT [16], adeclarative language, is intended for presenting specifications of binary data formats. Its syntax allows one expressing the specifications in a neat and well understandable form. FLEXT is accompanying by a code generator that can produce data reading source code in the imperative languages: PASCAL and C++. Now, it implements the code generation for the most widely used data types, but some complex types like that used in specifications of machine instruction encoding are not supported yet. KAITAI STRUCT [17] toolkit suggests a declarative language for describing formats of binary data and network packets. The language is primarily designed to describe communication protocols and container data formats. Specifications presented in this language can be translated into a source code for reading files in the one of the supported programming languages: C++, JAVA, JAVASCRIPT, PERL, PHP, PYTHON, RUBY, and GO. The toolkit implements a set of standard methods, which implement reading data from the stream in accordance with the type from the description. Its approach to code generation is similar to DATASCRIPT: all the data from the stream are sequentially loaded into the fields of the data structures.

Among the tools listed above, FLEXT and KAITAI STRUCT are the most suitable to be used in our case. Both provide the declarative languages for presenting file format specifications. Similarly, they consider a specification as a set of data type definitions. They support bit-oriented data (bit fields) and variant blocks. Both allow one generating source code of reading libraries for the raw data formats from the specifications. FLEXT language is more expressive, but KAITAI STRUCT language is based on well-known format, namely YAML. Moreover, KAITAI STRUCT supports more programming languages for the source code generation. We used both of them, for formally describing the raw data formats of TAIGA experiments. As a result, we generated reading libraries for each file format in the widespread programming languages including C/C++, JAVA and PYTHON.

## 3. Using Binary File Formats Specifications for Astroparticle Experiments

The considered raw data are generated and transmitted as packages by the facilities using the TCP protocol. Their file formats implement containers with simple structure. The developed format specifications consequentially match byte streams against data structures interpreting them. First, a specification optionally introduces format metadata and then it defines the section of data types and the data section containing definitions of variables.

A specification expressed in FLEXT consist of some definition blocks. The main blocks are the following: **const** defines constants that are results of calculations with variables of data blocks; **type** contains data type definitions; **data** is the data block with definitions of variables; **code** presents code block including address (shifts from the beginning of a file) and names of code parts. Fig. 1 shows the simplified specification of TUNKA-133 format specification expressed in FLEXT language: (*a*) — **unit**, a type for displaying time as two digits of integer; (*b*) — **THeader**, a header of data package; (*c*) — **TCompTime**, a computer time of a registered event; (*d*) — **TLinkData**, a data container using the defined types (**TCompTime**, **THeader**); (*e*) — **data** section specifies that data are an array of **TLinkData** structures.

A specification presented with KAITAI STRUCT includes the following blocks: **meta** — metadata (section describing format name, version, file extension); **doc** — a description of regular fields; **seq** sequentially lists definitions of variables used in the described format; **instances** — a description of fields that require additional processing; **enums** matches integer constants and some names; **types** contains user-defined data types. Fig. 2 demonstrates the auto-generated diagram for TUNKA-133 file format specification presented in KAITAI STRUCT. It shows the defined sections of the specification and the relations between their definitions.

```
                                                         ⎧ TCompTime struc
                  type                                   ⎪   uint h
a  { uint num+(2):displ=(Int(@))                       c ⎨   uint m
                                                         ⎪   uint s
   ⎧ THeader struc                                       ⎪   uint ms
   ⎪   word  Sign                                        ⎩ ends
   ⎪   byte  trSign
   ⎪   byte  errCnt                                      ⎧ TLinkData struc
   ⎪   word  N                                           ⎪   array[33]of THeader Packages
b  ⎨   byte  queryN                                    d ⎨   TCompTime T
   ⎪   ulong eventNum                                    ⎪   uint opticLen
   ⎪   ulong VME                                         ⎪   ulong cluEventNum
   ⎪   array[@.N-9]of byte Data                          ⎩ ends
   ⎪   byte  cluN
   ⎩ ends:assert[@.Sign=0xFFFF,@.trSign=0xA0]         e { data
                                                         0 array of TLinkData:[@:Size=FileSize] Hdr
```

Fig. 1. Simplified specification of TUNKA-133 file format expressed in FLEXT language.

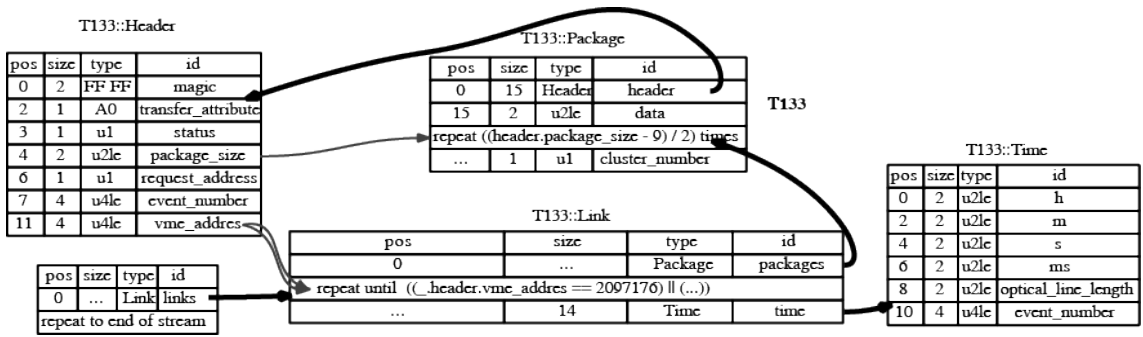

Fig. 2. Auto-generated diagram for TUNKA-133 file format specification presented in KAITAI STRUCT.

We implemented the specifications for all file formats of the considered experiments in both FLEXT and KAITAI STRUCT languages. It allowed us generating automatically source code of program libraries represented in the target programming languages (C++, JAVA, PYTHON, etc.) for parsing and verifying the raw experiment data.

## 6. Conclusion

The best practices of scientific data maintenance recommend keeping raw (unprocessed) data [18-19]. This allows facilitating future re-analysis with some improved analytical and data processing techniques as well as analytical reproducibility of published results. Astroparticle experiments involve with accumulating and processing a big volume of raw data. Each experiment uses some specific file formats for representing raw data. The considered case with raw data TAIGA experiment showed that such file formats could be insufficiently and weakly documented. Only a few experts involved in the experiments can interpret these data.

The paper considers our research experience on describing formally the binary file formats used in TAIGA experiment. We used FLEXT and KAITAI STRUCT toolsets to specify, parse and verify raw data of these formats. The implemented format specifications allowed us to generate source code for parsing and verifying the raw binary data in the target languages. The libraries were evaluated on real data. TUNKA-133, TUNKA-GRANDE, and TUNKA-REX formats were tested on about 89K files.Thanks to these libraries, we found that about 1.2% of these files contain corrupted data.. TAIGA-IACT and TUNKA-HISCORE formats were tested on about 120K files and 0.6% of them contains corrupted data. We suggest to use this approach for exchanging of astroparticle raw data. They can also simplify the software development for data aggregation from various sources in the case of multi-messenger analysis. We plan to share our experience of exporting raw data with other scientific collaborations.

# Acknowledgments

This work was financially supported by the Russian Scientific Foundation (Grant No 18-41-06003).